\pdfoutput=1

\documentclass[conference]{IEEEtran}
\IEEEoverridecommandlockouts
\usepackage{cite}
\usepackage{amsmath,amssymb,amsfonts}
\usepackage{algorithmic}
\usepackage{graphicx}
\graphicspath{{img/}}
\DeclareGraphicsExtensions{.pdf,.jpeg,.png}
\usepackage{textcomp}
\usepackage{xcolor}
\usepackage{soul}
\usepackage{float}
\usepackage{subfig}
\usepackage{xurl}
\usepackage{booktabs}
\usepackage{geometry}
\usepackage[activate={true,nocompatibility},final,tracking=true,kerning=true,spacing=true,factor=1100,stretch=10,shrink=10]{microtype}
\def\BibTeX{{\rm B\kern-.05em{\sc i\kern-.025em b}\kern-.08em
    T\kern-.1667em\lower.7ex\hbox{E}\kern-.125emX}}
\begin{document}

\title{Engineering and Experimentally Benchmarking a Serverless Edge Computing System\\
}

\author{\IEEEauthorblockN{Francisco~Carpio,~Marc~Michalke~and~Admela~Jukan}
    \IEEEauthorblockA{\textit{Institute of Computer and Network Engineering} \\
        \textit{Technische Universit{\"a}t Braunschweig}, Germany \\
        E-mail: \{f.carpio, m.michalke, a.jukan\}@tu-bs.de}
}

\maketitle

\begin{abstract}
    Thanks to the latest advances in containerization, the serverless edge
    computing model is becoming close to reality. Serverless at the edge is
    expected to enable low latency applications with fast autoscaling
    mechanisms, all running on heterogeneous and resource-constrained devices.
    In this work, we engineer and experimentally benchmark a serverless edge
    computing system architecture. We deploy a decentralized edge computing
    platform for serverless applications providing processing, storage, and
    communication capabilities using only open-source software, running over
    heterogeneous resources (e.g., virtual machines, Raspberry Pis, or bare
    metal servers, etc). To achieve that, we provision an overlay-network based
    on Nebula network agnostic technology, running over private or public
    networks, and use K3s to provide hardware abstraction. We benchmark the
    system in terms of response times, throughput and scalability using
    different hardware devices connected through the public Internet. The
    results show that while serverless is feasible on heterogeneous devices
    showing a good performance on constrained devices, such as Raspberry Pis,
    the lack of support when determining computational power and network
    characterization leaves much room for improvement in edge environments.
\end{abstract}


\section{Introduction}

With the improvement of computing capabilities and network bandwidth for
end-users, computing at the edge is gaining popularity when serving
latency-sensitive applications while also reducing communication to remote data
centers. Also, cloud providers are offering solutions that include edge
infrastructures by deploying on-premises nodes to provide low-latency and
data-locality services. This evolution has been greatly supported by
containerization, an enabling technology that can address the resource
heterogeneity at the edge. Known container orchestrators, such as Kubernetes
have been widely proven to provide service resilience against hardware and
container failures and reducing deployment and maintenance costs by providing
seamless autoscaling to meet demands for computing, when and where needed.
Containerization has also brought a huge advantage to DevOps by facilitating
development, testing, and monitoring tasks and eliminating conflicts that
applications can present when running in different environments.

Today, research in serverless edge computing is primarily focused to either
solving new theoretical problems introduced by, for instance, the volatility of
edge devices, or to benchmarking specific software tools, such as container
runtimes or serverless platforms, or yet to benchmarking different types of
hardware devices. What is currently missing are the benchmarking studies of the
entire edge computing systems, engineered in real-world scenarios.  Such studies
are rather critical when building systems to work in decentralized environments
at the edge, running over heterogeneous, constrained, and networked devices. For
instance, limitations on the capabilities of edge devices can make some exiting
tools fundamentally unsuitable due to processor architectures incompatibility
(e.g., AMD64, ARMv7, ARMv8, etc.). While many cloud computing tools are
open-source and already in production, there is still a lack of community-driven
projects engineering and benchmarking edge computing systems with these tools.

We engineer and benchmark a serverless edge computing system based on
open-source software tools running over different network domains. We use Nebula
to provide network abstractions, while connecting devices in different sites,
and K3s, a lightweight distribution of Kubernetes designed for constrained
devices, as container orchestrator to provide hardware abstractions. To provide
processing, communication, and storage capabilities, we make use of
lightweight and production-ready tools, including OpenFaaS, NATS, and
Elasticsearch, respectively. We first share the lessons learnt in the design of
the system including communication, storage, and processing capabilities. We
then benchmark the system in terms of response time, throughput and scalability.
The results show that serverless edge computing is feasible on heterogeneous
devices, showing good performance on constrained devices, such as Raspberry
Pis. Tools are still missing in support of determining computational power and
to monitoring the network state in container orchestrators, and require further
improvements when engineering edge computing systems.

The rest of the paper is organized as follows. Section II presents related work.
Section III describes the architectural design. Section IV analyzes the
performance of the system under different scenarios and Section V concludes the
paper.

\begin{figure*}[!t]
    \centering
    \includegraphics[width=0.9\textwidth]{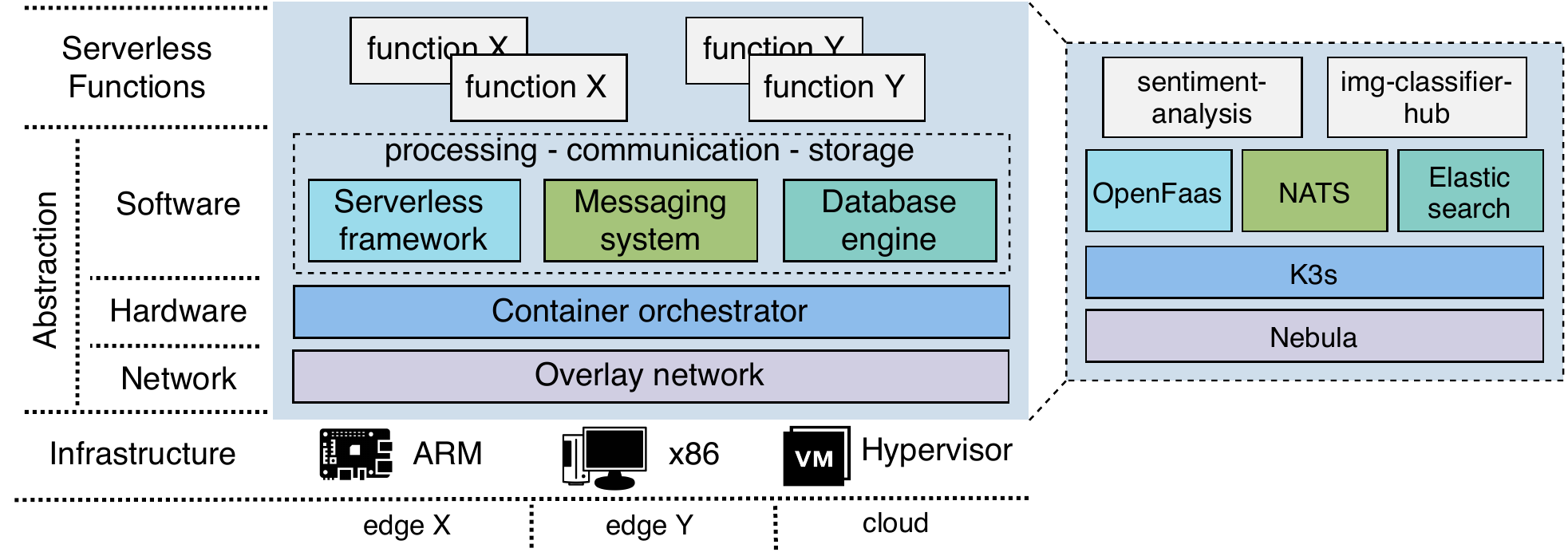}
    \caption{Architectural design of the proposed edge computing system model}
    \label{fig:arch}
\end{figure*}

\section{Related Work}



While the traditional Platform-as-a-Service (PaaS) model hides the backend from
customers when facilitating development, deployment, and management tasks,
system scaling remains a challenge for users \cite{VanEyk2018}. In contrast, the
Function-as-a-Service (FaaS) model removes decision making regarding scaling
thresholds, reduces costs by charging only when applications are triggered, and
reduces starting times at expenses of having less control over the deployment
environment \cite{Castro2019}. This transition, however, would not have been
possible without the latest advances in containerization, especially in edge
devices \cite{Morabito2017}. In addition, the high performance that container
orchestrators, such as Kubernetes, have already been proven in the cloud
context, has been improving in the edge context by replacing, for instance,
schedulers that consider the state of the network to address latency
\cite{Santos2019}. Unlike what is known for virtual machines (VMs) where
resources are usually underused due to overprovisioning techniques necessary to
handle peaks of requests, in serverless computing, resources can be provisioned
instantly on demand while minimizing the costs and efforts for developers when
dealing with scalability \cite{Adzic2017}.



Since traditional centralized cloud infrastructures are not expected to be able
to handle the vast amount of data generated by IoT devices while assuring
real-time interaction \cite{Carpio2020}, the idea of having a network of
computing devices closer to the end-users becomes essential \cite{Li2017a}. Edge
computing is expected to be able to deal with high intensive computing
applications, such as deep learning models \cite{Wang2020}, while handling a
high number of requests and ensuring very low latency \cite{Das2019a}. Despite
the advantages, the edge introduces significant challenges, including resource
capability constraints, heterogeneity, resiliency and security, which still need
to be addressed \cite{Caprolu2019}. In this area, containerization becomes even
more essential to address device heterogeneity \cite{Morabito2018}. The adoption
of both serverless functions and containerization techniques, is the natural
next step forward in the edge computing evolution, as also demonstrated in the
commercial context \cite{Das2019a}.



Despite the huge attraction that edge computing is presenting to both academia
and industry, benchmarking studies of open source serverless platforms in the
edge context are few and far between. Most of the related work analyzes and
benchmarks the performance of cloud tools \cite{Loghin2019}, or compares various
serverless frameworks \cite{Palade2019} in the edge context. To further
contribute to this evolution, we propose the design of a serverless edge-centric
computing system using open source tools already used in production and
benchmark the system.

\section{Architectural Design}\label{sec:arch}

We set the goal of building a system that can provide abstractions of
processing, storage, and communication resources, so that dependency conflicts
and redundancies are avoided, which is a challenge that developers and system
administrators need to face. The proposed system architecture is illustrated in
Fig. \ref{fig:arch}. First, it should be noted that the proposed architecture is
built considering different network domains (i.e. private/public networks or
clouds) and with the requirement set that it should be deployable over
bare-metal (ARM- or x86-based) or hypervisor-based VMs (Infrastructure). The
system abstraction includes three categories: network, hardware, and software.
Network abstraction is achieved by creating an overlay peer-to-peer network for
the interconnection of the devices such that firewalls are avoided. Hardware
abstraction is achieved by using of a container orchestrator which abstracts all
the devices, deals with container lifecycle and scaling. Software abstraction is
achieved by providing processing, communication, and storage capabilities, using
a serverless framework, messaging system, and a database engine, as services to
be used by applications. These applications can be created by deploying
serverless functions (function x, function Y) which are agnostic from the
underlying abstraction. For the reminder of this section, we go into
implementation details of the abstraction layers, as well as the serverless
functions used for the benchmarking of the system.

\subsection{Network abstraction: Overlay networks}

When connecting devices located behind NATs (Network Address Translation) or
firewalls, it is required to either apply manual NAT configurations, firewall
rules, or to use traditional privacy-oriented VPN solutions (e.g., OpenVPN or
Wireguard). While these VPN solutions can create mesh VPN networks by tunneling
all traffic, their configuration is laborious and does not scale with an increased
number of connected devices. Instead, new tools such as Nebula or Zerotier make
use of easily scalable beaconing servers and UDP hole punching technique to
directly interconnect devices over the network avoiding manual firewall
configurations and creating, in this way, mesh overlay networks. These tools are
not privacy-oriented, so not all the traffic generated by a host is tunneled,
but only additional network interfaces are created and used when communicating
with the nodes within the created overlay network. Since this solution is simpler
and more scalable in the edge context, in our implementation we choose Nebula,
which is free and open-source developed by Slack's team. Nebula makes use of
mutual authentication using certificates, traffic encryption via Diffie-Hellman
key exchange, and AES-256-GCM in its default configuration, as well as traffic
filtering. The clients connect to the lighthouse (i.e., beaconing server) first
and then start the negotiation of paths between the clients. These paths remain
configured by both clients using the UDP connections for keep-alive messages, if
no data is transferred.

\subsection{Hardware abstraction: Container orchestration}

Container orchestrators are responsible for integrating, scaling, and managing
containers, while at the same time providing security, networking, service
discovery, monitoring, etc. Some cloud providers, such as Amazon, have developed
their own orchestrators, while others support multiple, such as Kubernetes,
Apache Mesos Marathon, or Docker Swarm. From all available options, Kubernetes
is the most widely adopted in large production environments. It has the
advantage of not being constrained to one specific container runtime, but open
to Docker, containerd, CRI-O, or any following the Kubernetes Container Runtime
Interface specifications. We choose a lightweight Kubernetes distribution built
for the edge, known as K3s, which comes with all needed components into a single
binary of less than 40MB. K3s achieves this small size compared to other
distributions by dropping some storage drivers and cloud providers not needed
for its purpose.

\subsection{Software abstraction: Serverless framework}

While containerization facilitates the packaging and deployment of applications,
administrators are still responsible for the scalability management of container
orchestration systems. The serverless model removes this need for the
administrator to manage the scaling. Serverless functions are small pieces of
code that are only executed when they are explicitly triggered. Typically these
functions have a specific purpose, are stateless, and run for short periods of
time. Some cloud providers, such as Amazon with Lambda or Google with Cloud
functions, are already offering the deployment of functions by providing IDEs,
SDKs, plugins, etc. In the open-source community, serverless platforms such as
OpenFaaS, OpenWhisk, or Kubeless are currently under development. In our
implementation, we make use of OpenFaaS due to the maturity and minimum hardware
requirements compared to the other options. OpenFaaS provides a means to deploy
containerized functions without having to deal with context-specific APIs.

\subsection{Software abstraction: Messaging system}

When developing applications based on microservice architectures, the design of
APIs for the communication between components can follow different styles such
as RPC, REST, query, or event-driven, where their choice will depend on the use
case. When developing a scalable system for a larger number of devices,
client-server communication is not feasible; instead, the publish-subscribe
model is preferable. In this context, while traditional publish-subscribe
message brokers (e.g., RabbitMQ or AMQP) can serve that purpose, event-driven
messaging systems, such as Apache Kafka or NATS bring some advantages. These
messaging systems are typically based on distributed logging which offers good
scalability, persistency, high availability, and fault-tolerance while assuring
real-time communication. Their persistency allows microservices to restore their
state anytime in case of failures, which is critical in large-scale cloud
applications, but also the edge benefits from this feature where systems tend to
be less reliable in terms of hardware and communication. From the available
open-source platforms, we choose NATS which is lightweight compared to other
options and is officially available via Docker Hub.

\subsection{Software abstraction: Database engine}

Containerization introduces a new issue regarding data persistency, which tools
like Docker or Kubernetes solve by making use of volumes to persist data on disk
that can be restored when containers restart. Containerization allows for multiple
database instances with subsets of the whole data instead of traditional
monolithic databases. While database management systems can be either relational
or non-relational, the choice strictly depends on the kind of data. Since, in
our case, no data schemas are predefined, NoSQL is more adequate. Of this type,
there are four main categories: key-value store, document-, graph-, and
column-based. Document-based are indeed those designed for flexibility and
typically make use of XML or JSON formats to store data which enables developers
to perform integration with their code without redundant conversions. While
there are multiple tools making use of this type of database as a backend
system, we focus on search engines, which are optimized for searching and can
deal with large amounts of content. The most popular open-source option and
widely used in the industry is Elasticsearch; a scalable, distributed
full-text search engine that makes use of JSON documents and offers a built-in
RESTful API.

\subsection{Serverless Functions}

We use two different serverless functions for the testing of the proposed
system; \emph{sentiment-analysis} and \emph{img-classifier-hub}. The functions
are written in Python and containerized following the guidelines provided by
OpenFaas to be production-ready. The \emph{sentiment-analysis} makes use of the
TextBlob\cite{loria_2020_sloriatextblob} project to analyze the polarity and
subjectivity of a given text. The \emph{img-classifier-hub} is based on machine
learning, making use of the Inception v3 model \cite{Szegedy_2016} using
Tensorflow Hub to classify images based on a set of predefined labels. Both
functions have been containerized for both AMD64 and ARM64 CPU
architectures\footnote{Carpio, Francisco. Fcarp10/Openfaas-Functions. GitHub, 23
    Apr. 2021, github.com/fcarp10/openfaas-functions. Accessed 28 Apr. 2021.} and
are publicly available via Docker Hub.

\section{Experimental Benchmarking}

\begin{figure}[!t]
    \centering
    \includegraphics[width=1.0\columnwidth]{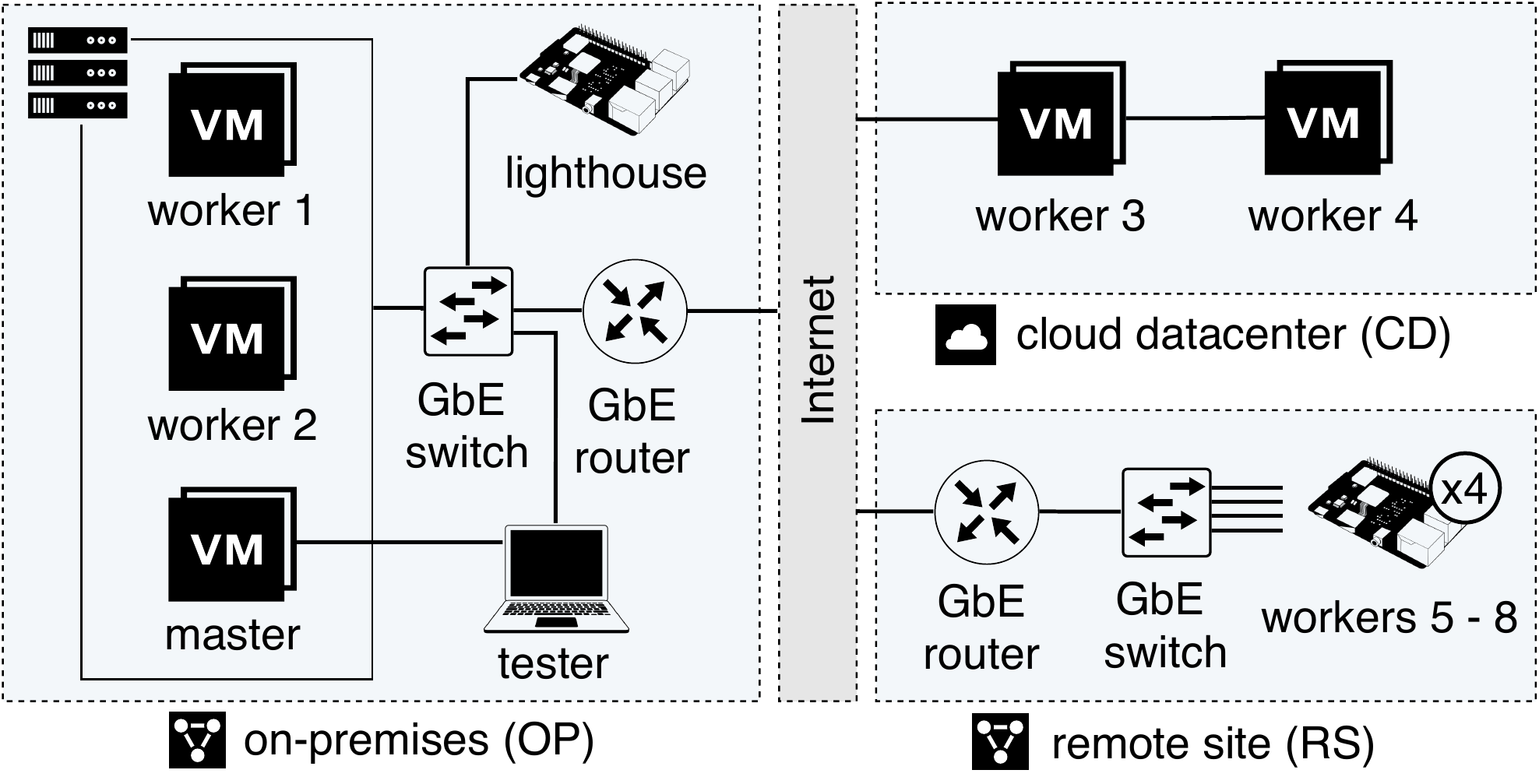}
    \caption{Testbed}
    \label{fig:testbed}
\end{figure}

\begin{table}[t!]
    \centering
    \caption{Device specifications}
    \label{tab:device-specs}
    \begin{tabular}{lcccc}
        \toprule
        \textbf{Name} & \textbf{Type} & \textbf{Location} & \textbf{CPU}  & \textbf{Memory} \\
        \midrule
        Hypervisor    & workstation   & OP                & Ryzen 9 5900X & 32 GB           \\
        Master        & VM            & OP                & 4 vCPU        & 8 GB            \\
        Workers 1-2   & VM            & OP                & 2 vCPU        & 8 GB            \\
        Tester        & laptop        & OP                & i7-3632QM     & 12 GB           \\
        Lighthouse    & RPi 4         & OP                & Cortex-A72    & 8 GB            \\
        RPi workers   & RPi 4         & RS                & Cortex-A72    & 8 GB            \\
        Workers 3-4   & VM            & CD                & 2 vCPU        & 8 GB            \\
        \bottomrule
    \end{tabular}
\end{table}

The experimental benchmarking and performance evaluation is carried out in the
testbed implementation shown in Fig. \ref{fig:testbed}. The testbed connects to
three different sites via Nebula: on-premises (Springe, Germany), a remote site
(Braunschweig, Germany) and a cloud datacenter (Brasilia, Brasil). On-premises
consists of a workstation hosting 3 VMs (one master and 2 workers), a Raspberry
Pi (RPi) running as lighthouse and a laptop as tester machine with one GbE
interface connected directly to an interface of the workstation and a second GbE
interface connected to a GbE switch, which is at the same time connected to a
GbE router with 10/200 Mbps (uplink/downlink) internet connection. The remote
site consists of 4 RPis working as worker nodes connected to a GbE switch and a
GbE router with a 0.5/1 Gbps internet connection. In the remote datacenter, 2
VMs as worker nodes are allocated by the GigaCandanga datacenter. The remaining
hardware specifications are described in Table \ref{tab:device-specs}. K3s is
then deployed on the master and worker nodes using the IP addresses provided by
Nebula. Once the cluster is set up, the OpenFaaS components are deployed and
constrained to the master node while the functions are deployable on any worker.
NATS and Elasticsearch are deployed on every worker in the cluster, so each
component has one replica per worker node. For the evaluation, four different
scenarios are considered depending on the resources: 1) on-premises only (OP),
2) remote site only (RS), 3) cloud datacenter only (CD) and 4) all sites
together (AS). For all scenarios, benchmarking is performed by redeploying the
cluster and then running the same series of tests from the tester machine to
ensure comparability.

\subsection{Networking}

The network latencies are measured over TCP connections using
\emph{tcp-latency}\footnote{González, Daniel. Dgzlopes/Tcp-Latency. GitHub, 26
    Apr. 2021, github.com/dgzlopes/tcp-latency. Accessed 26 Apr. 2021.}, a python
tool based on \emph{netperf}, over 500 repetitions (see Table
\ref{tab:latency}). As we can expect, the obtained speeds when using Nebula are
only slightly slower compared to baremetal in all the cases when testing nodes
within the same site (i.e., from RS to RS, from CD to CD, etc.). This is due to
Nebula nodes only asking the lighthouse for the target nodes location once and
then following the shortest path for the traffic; in this case, Nebula traffic
never leaves the site. More interesting is the case when Nebula speeds are even
faster than baremetal (i.e., from RS to test and CD to test). This behavior is
due to Nebula encapsulating the traffic using UDP which suggests traffic
prioritization by intermediate routers in the network since devices located
within the same site have been confirmed to not exhibit this behavior. This
observation is not unique to our setup but has been confirmed by Nebula
developers to also occur when benchmarking their systems.

\begin{table}[t!]
    \centering
    \caption{Baremetal/Nebula latencies between devices (ms)}\label{tab:latency}
    \begin{tabular}{lcccc}
        \toprule
        \textbf{from - to} & \textbf{mean} & \textbf{min} & \textbf{max} & \textbf{std} \\
        \midrule
        RS - RS            & 0.21/1.23     & 0.19/0.85    & 0.30/1.77    & 0.02/0.28    \\
        RS - test          & 32.57/27.57   & 28.37/25.74  & 44.44/34.33  & 2.63/1.63    \\
        CD - CD            & 0.85/1.32     & 0.31/0.46    & 4.24/10.06   & 0.55/1.13    \\
        CD - test          & 238.9/231.5   & 227.7/229.1  & 451.9/242.4  & 30.35/1.75   \\
        OP - OP            & 0.33/0.78     & 0.24/0.57    & 0.46/1.25    & 0.05/0.1     \\
        OP - test          & 0.57/1.17     & 0.42/0.79    & 0.69/1.97    & 0.05/0.17    \\
        RS - CD            & 231.0/232.1   & 230.5/231.6  & 231.6/234.2  & 0.24/0.28    \\
        \bottomrule
    \end{tabular}
\end{table}

\subsection{Processing}

\begin{figure*}[!t]
    \centering
    \subfloat[Response time of \emph{sentiment-analysis} function]{\includegraphics[width=0.5\textwidth]{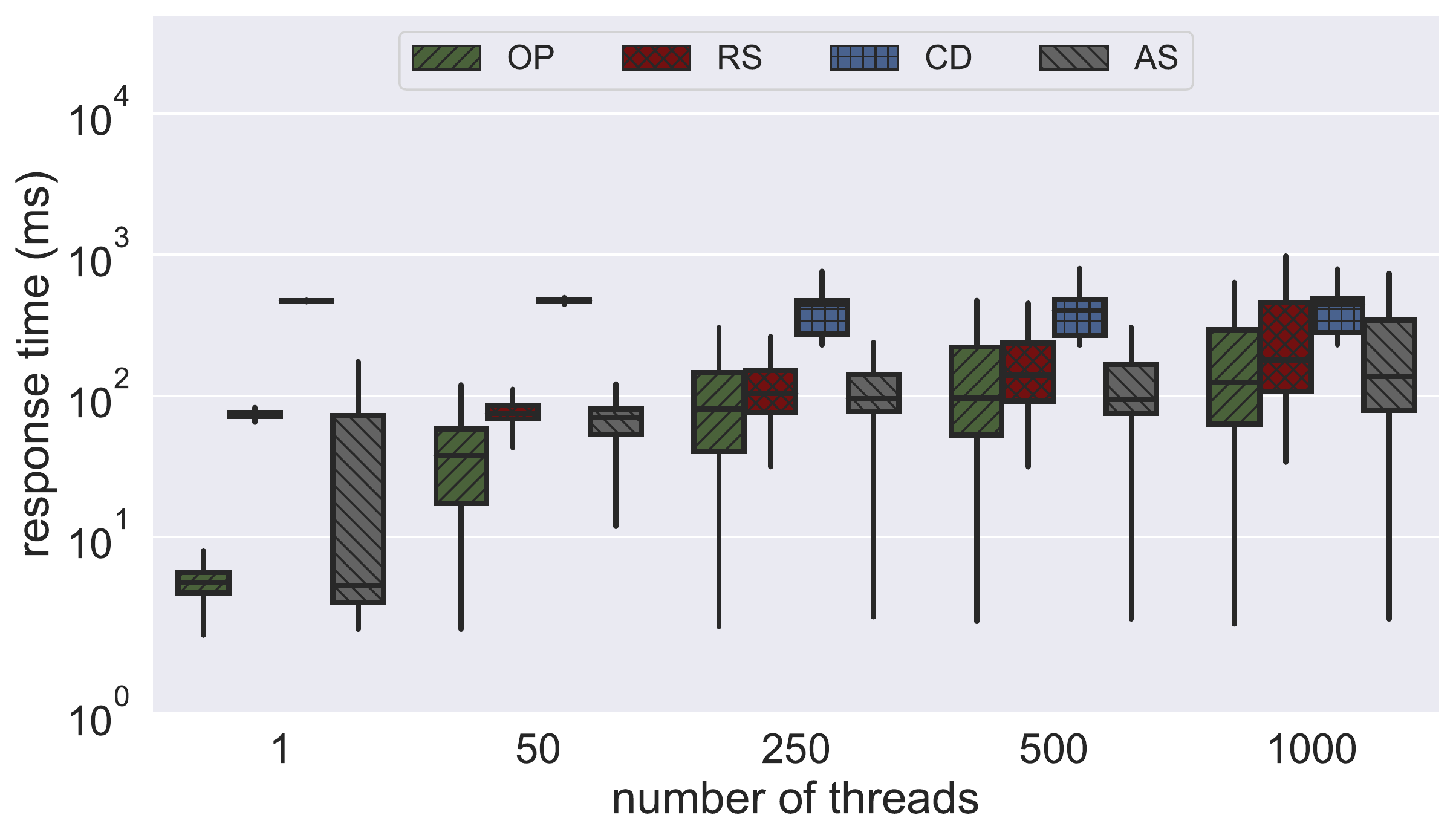}\label{openfaas-sentiment}}
    \hfil \hfil \hfil
    \subfloat[Response time of \emph{img-classifier-hub} function]
    {\includegraphics[width=0.5\textwidth]{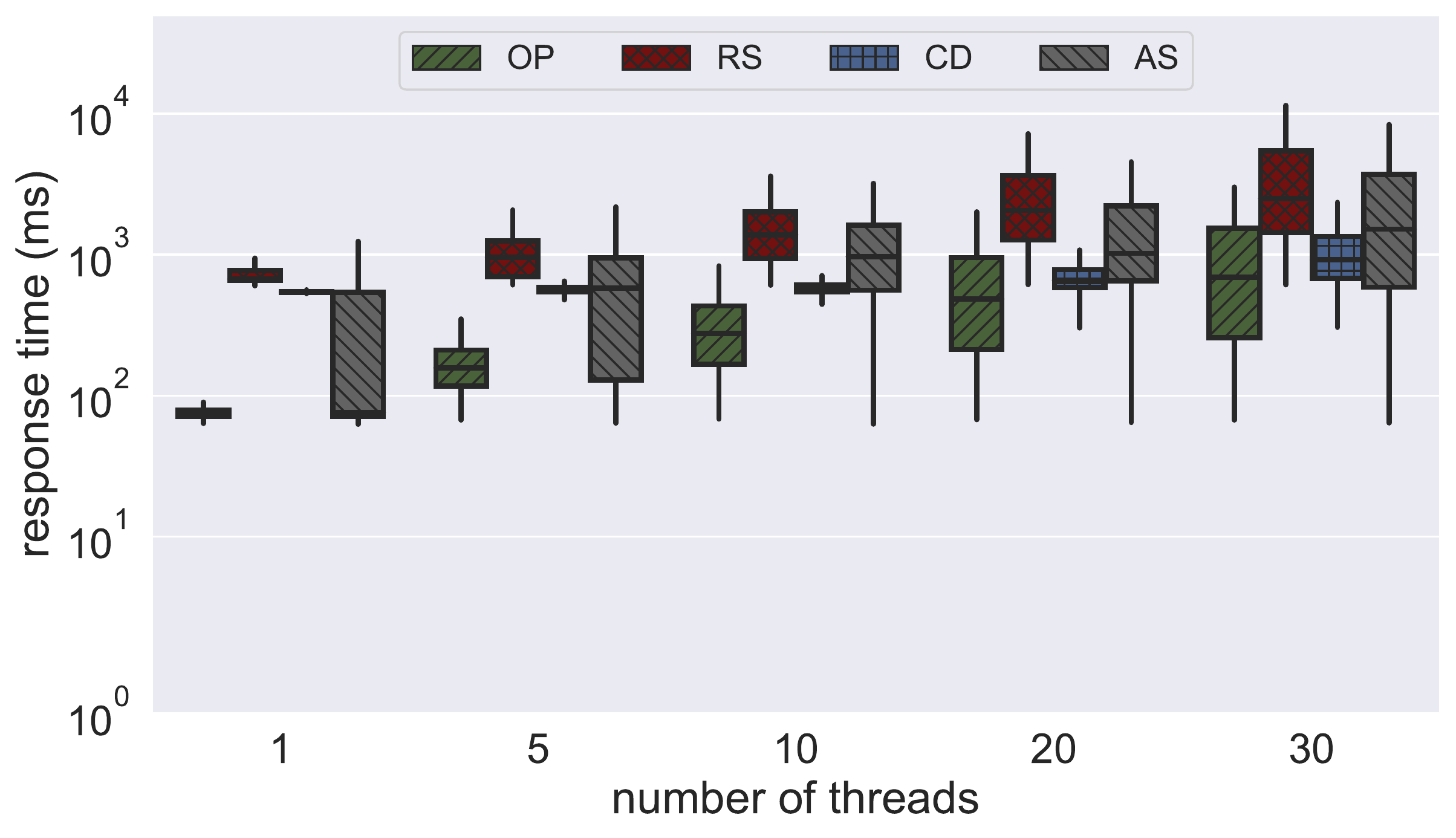}\label{openfaas-inception}}
    \caption{\emph{sentiment} and \emph{img-classifier-hub} average response times}
\end{figure*}

We use \emph{Hey}\footnote{Dogan, Jaana. Rakyll/Hey. GitHub, 15 Apr. 2021,
    github.com/rakyll/hey. } as one of the third-party tools recommended by the
OpenFaaS project, to benchmark the OpenFaaS functions. To trigger the functions,
HTTP POST requests are sent with a specific sentence to analyze, in the case of
\emph{sentiment-analysis}, or with the URL of an image, for the case of
\emph{img-classifier-hub}. The functions are stressed with a concurrent number
of threads, from 1 to 1000 for \emph{sentiment-analysis} and from 1 to 30 for
\emph{img-classifier-hub}, where the total number of requests is constant for
all tests, being 200000 for \emph{sentiment-analysis} and 25000 for
\emph{img-classifier-hub}. After each test, the system is released for 15
minutes to scale down. For all tests, \emph{-disable-compression} and
\emph{-disable-keepalive} parameters are provided to ensure that the TCP
sessions are not reused and the load is distributed amongst all available
containers.

\subsubsection{sentiment-analysis}
Fig. \ref{openfaas-sentiment} shows the response of \emph{sentiment-analysis}
with different numbers of concurrent threads using boxplots for the four
scenarios: OP, RS, CD, and AS. In the OP case, we can see how the median value
is almost always the lowest compared to the other cases, and the interquartile
range increases with the number of threads due to the number of replicas being
limited to the available resources. In the RS case, the median response times
are always higher than OP also with a lower interquartile range, basically due
to the effect of the network delay. This effect is more obvious in the CD case
where, despite the processing capabilities being the same as in OP, the huge
delay of the network makes the response times quite similar, independent of the
number of threads. So in this case the influence of the processing time is very
small compared to the time added by the network latency. In most tests, the AS
case behaves closer to the RS case unless stressed with a single thread. This is
due to Kubernetes typically spawning replicas based on the number of CPU cores
and the amount of RAM that the node has, so here the RPis are seen as the nodes
with the most resources (i.e., 4 cores and 8GB RAM per RPi), even though those
resources deliver lower effective performance than the ones of the VMs. This
leads to no benefit being provided by clustering all devices in this scenario
for this specific function.

\subsubsection{img-classifier-hub}
The response time obtained by the \emph{img-classifier-hub} is shown in Fig.
\ref{openfaas-inception}. Contrary to the previous case, this function is highly
CPU intensive which leads to higher overall response times. For this reason, the
maximum number of concurrent threads is limited to 30, assuring, in this way, a
low number of bad responses resulting from timeouts occurring when requests
cannot be served. For a fair comparison, only valid responses are considered.
With these functions, we can see how the processing time in the RS case has a
much higher impact on the response time than the network latency due to the RPis
being far slower in processing than the VMs. In this case, the VMs in the cloud
case exceed the RPis response speed even though the network latency is much
higher. For the AS case here, we can see how the interquartile range is spread
and the median is positioned between the other cases. This is again due to the
Kubernetes pods being assigned across different sites, so the expected response
times vary depending on which container the request is processed.

\subsection{Messaging}

\begin{figure}[!t]
    \centering
    \subfloat[Publishers - throughput]{\includegraphics[width=0.24\textwidth]{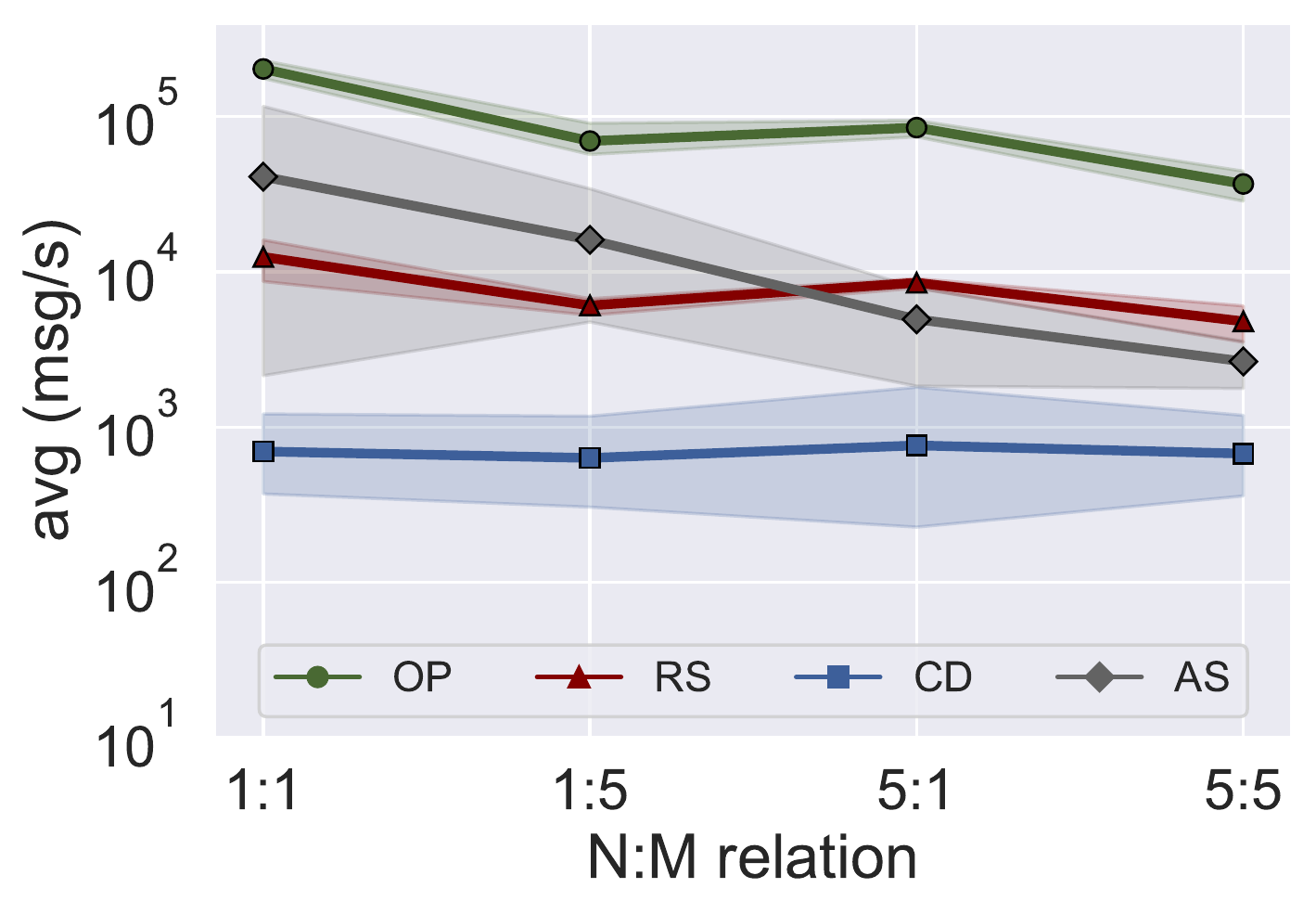}\label{nats-pubs}}
    \hfil
    \subfloat[Subscribers - throughput]{\includegraphics[width=0.24\textwidth]{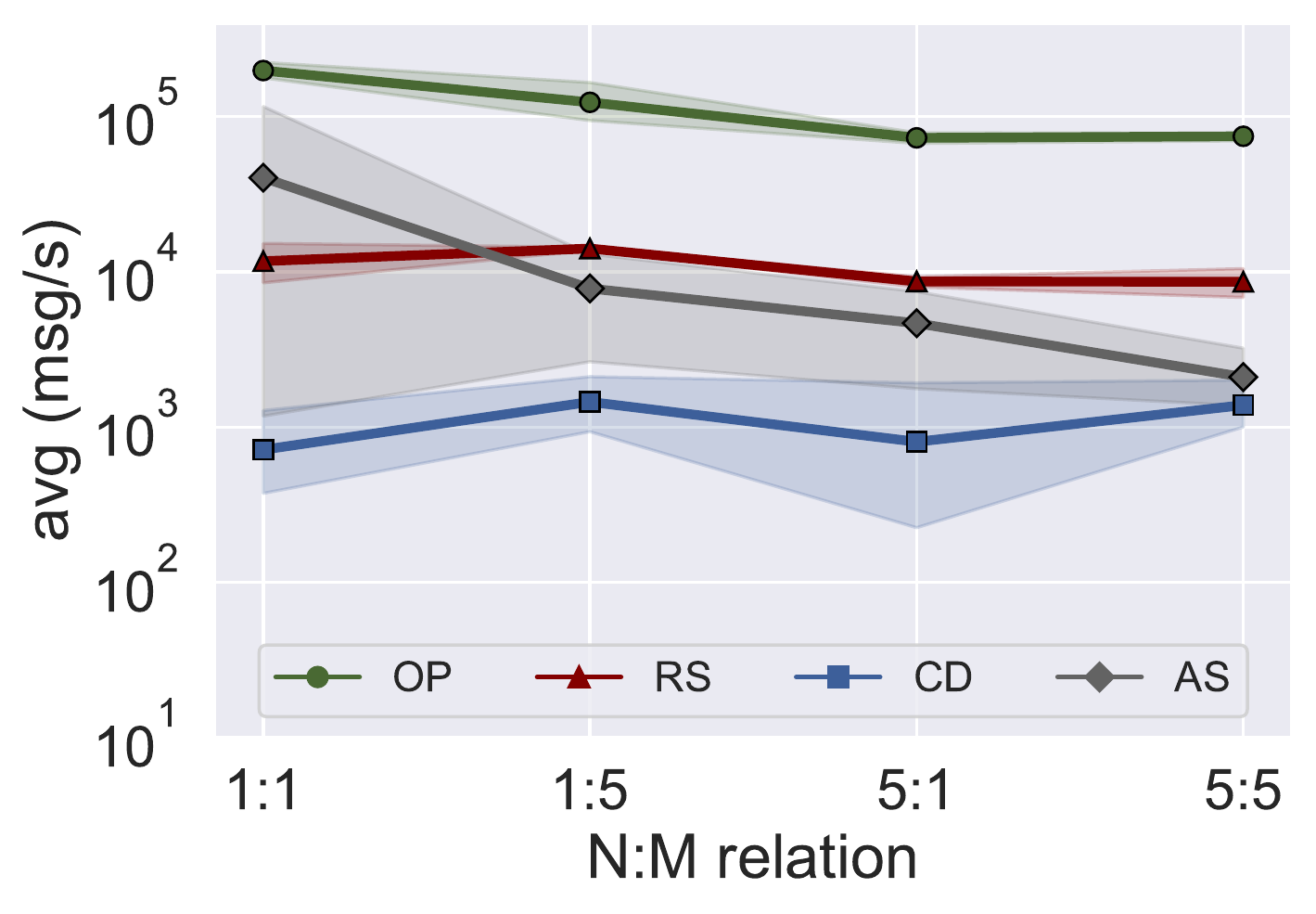}\label{nats-subs}}
    \caption{NATS performance}
\end{figure}

\emph{NATS-bench} is maintained by the NATS project and offers automatic
deployment of N to M number of publishers and subscribers, respectively, for
performance evaluation. The parameters for each test are 64 Bytes for message
size and 10000 for the total number of messages. The whole benchmark is repeated
five times and the resulting throughput for each test is averaged. Fig.
\ref{nats-pubs} and Fig. \ref{nats-subs} show the throughput of the publishers
and subscribers in messages per second, respectively comparing all four
scenarios for the configurations 1:1, 1:5, 5:1, and 5:5. As expected, the OP
case is the one offering a higher throughput compared to the other cases since
the node in use is closer to the tester machine, and the performance decreases
when adding more publishers/subscribers. The RS performs around 10 times worse
than the OP case, but here the degradation of performance with a higher number
of publishers/subscribers is not that as evident as in the previous case. This
is most likely since there are 4 different devices (i.e. RPis) in the RS case,
so 4 replicas of NATS instead of only 2 (one per VM) as we have in the OP. So
even though the RPis have slower CPUs, they can handle more
publishers/subscribers without degrading performance. The performance of the CD
case is, in this case, 10 times worse than RS with a considerable variance.
Since the computational resources are the same as in the OP case, this variance
is clearly introduced by the network latency, probably causing stream
interruptions. The AS case shows extreme variance in 1:1 due to the pods serving
the requests being assigned randomly to the different nodes in the cluster.
Also, every request requires a writing operation that triggers synchronization
between replicas, leading to an additional penalty even if a subscriber gets
assigned to the same instance serving the publisher.

\subsection{Storage}

\begin{figure}[!t]
    \centering
    \subfloat[Latency per scenario]{\label{rally-lat-scenarios}\includegraphics[width=0.24\textwidth]{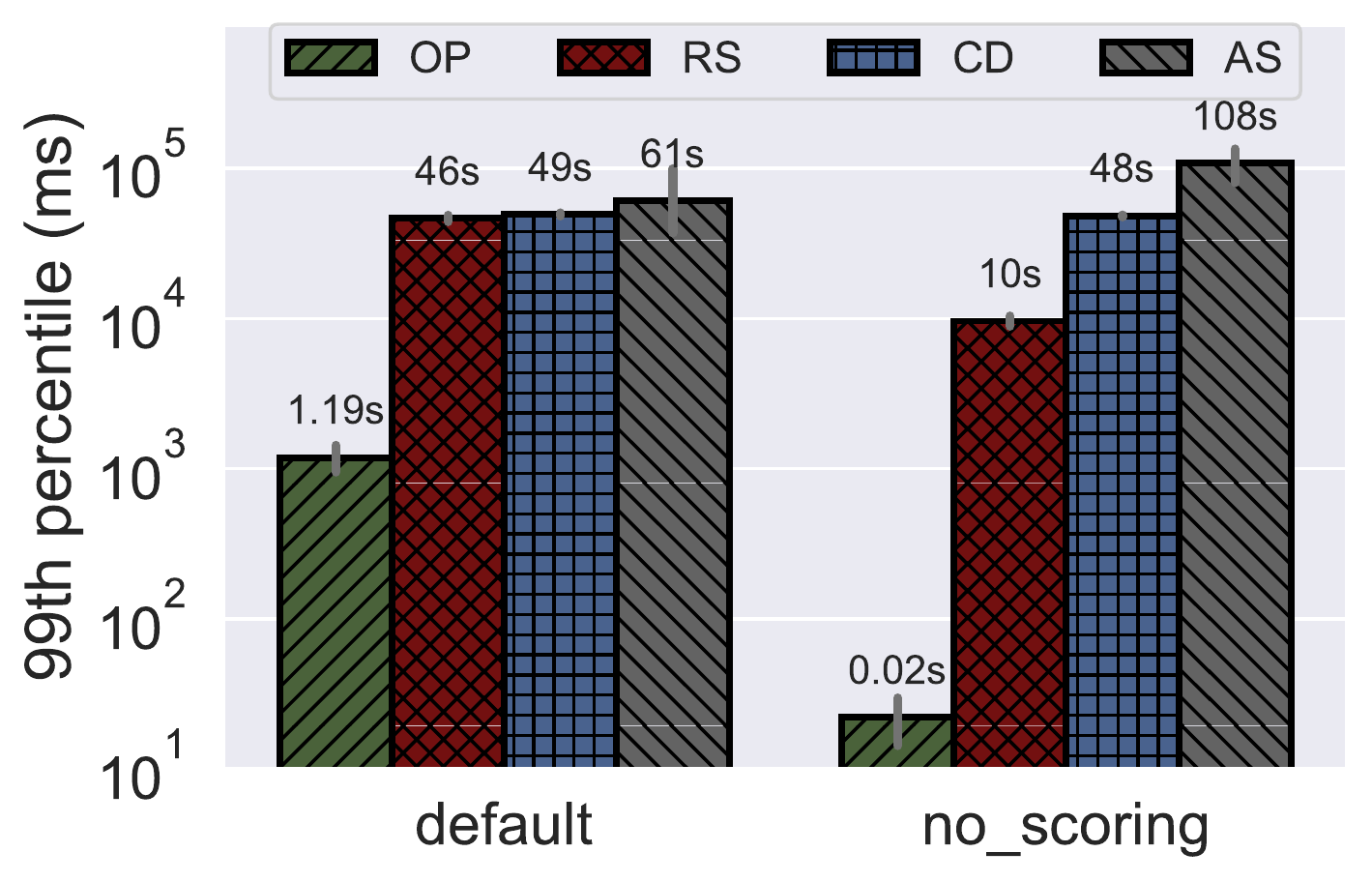}}
    \hfil
    \subfloat[Latency of \emph{all} scenario]{\label{rally-lat-all}\includegraphics[width=0.24\textwidth]{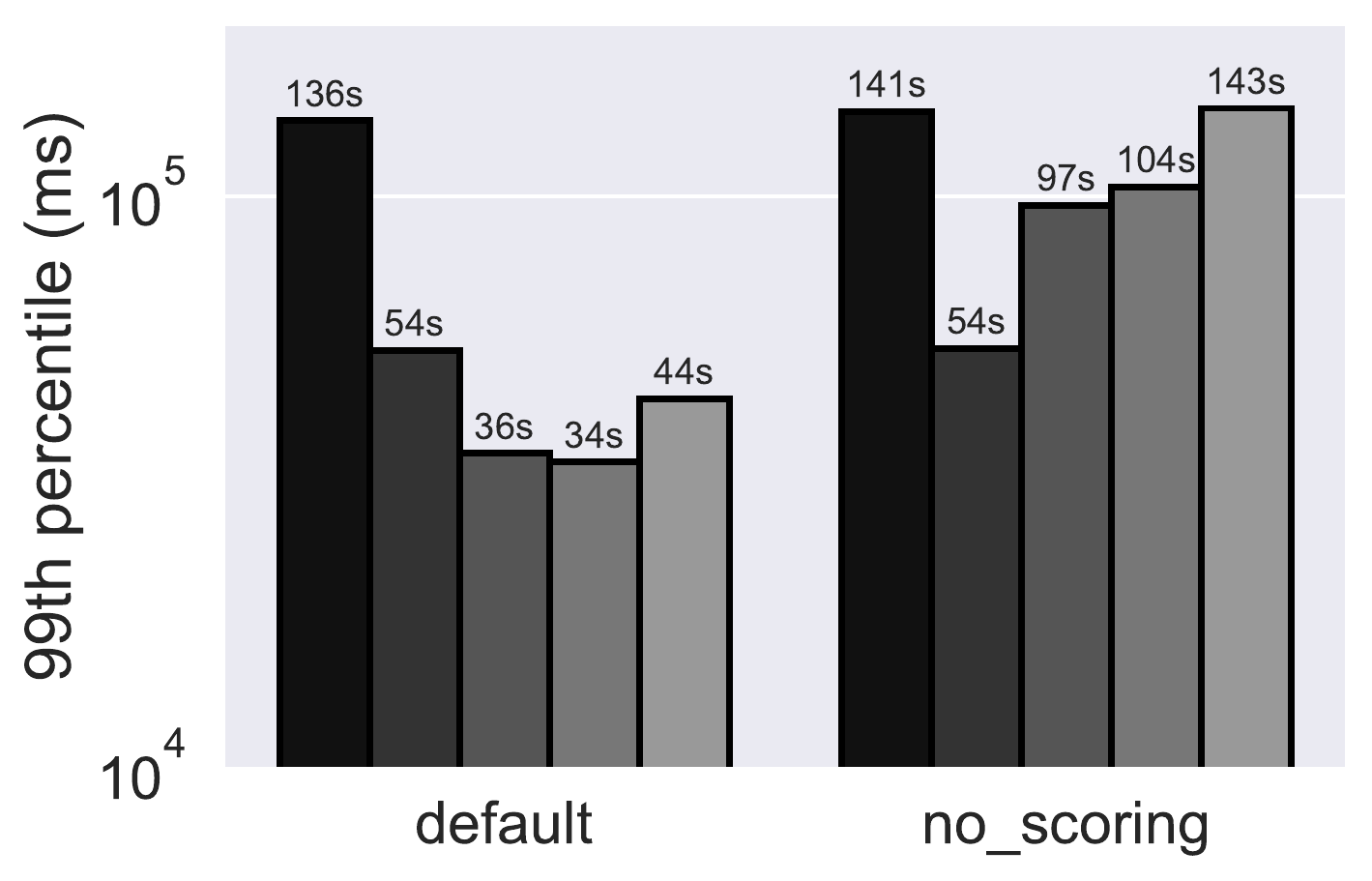}}
    \caption{Elasticsearch performance using percolator queries}
    \label{rally-latencies}
\end{figure}

\emph{Rally} is the official Elasticsearch benchmark tool that offers multiple
types of workloads (a.k.a., tracks) for testing. In our case, we run the
\emph{Percolator} test which performs reverse queries from the AOL search
dataset leak in 2006. The reason for selecting this test is its performing of
many queries of small size, which is a typical, expected scenario in edge
environments where they are supposed to deal with high amounts of data sources,
such as sensors, producing high numbers of small data records. By default, the
percolator benchmark is performed using scoring, which is a metric to determine
the relevance of the retrieved documents based on the queries. Scoring, however,
requires extra computational power since it ranks documents based on their
relevance. For this reason, Fig. \ref{rally-lat-scenarios} shows the latencies
for all four scenarios, when using scoring (by default) and when scoring is
deactivated, to see the impact on the results. We can see how the OP case
performs by far the best under both settings due to the superior combination of
low latency to the node and high computation power. In the RS case, however, we
can see the effect of scoring on the results due to the RPis limited
computational power compared to VMs. For this reason, the CD case performs
similarly when scoring is enabled and only when it is disabled, the RPis can
outperform the VMs in the cloud by a factor of around 4.8 on average, showing the
dramatic influence of the network latency. For the AS case, the performance is
affected by a high variance, represented with more details in Fig.
\ref{rally-lat-all}, due to the synchronization time between backup shards located
at different nodes, which in the worst case can perform worse than RS or CD
cases.

\section{Conclusions}

We engineered a decentralized networked computing platform for
serverless applications providing processing, storage, and communication
capabilities over heterogeneous devices. We benchmarked the system in terms of
response times, throughput and scalability. The results show how constrained
devices such as Raspberry Pis with a typical network latency can perform better
than VMs allocated in remote datacenters for most of the tasks when no excessive
computational power is required. Also, clustering heterogeneous devices across
the internet without optimizing the placement of containers results in
unpredictable performance. From the obtained results we can extract that the
adaptation of tools, such as Kubernetes, for heterogeneous edge devices still
needs to consider more hardware specifications for an optimal categorization of
computing power which differs much more compared to cloud environments. Also,
all tested tools in this work lack support for networking characterization which
causes a non-optimal selection of containers when serving requests.

\section*{Acknowledgment}

We would like to thank Professor André Drummond, from the University of
Brasilia, and GigaCandanga in Brasilia, Brazil, for making their cloud
infrastructure available for the implementation of the testbed.

\bibliographystyle{IEEEtran}
\bibliography{ref}

\begin{thebibliography}{10}
\providecommand{\url}[1]{#1}
\csname url@samestyle\endcsname
\providecommand{\newblock}{\relax}
\providecommand{\bibinfo}[2]{#2}
\providecommand{\BIBentrySTDinterwordspacing}{\spaceskip=0pt\relax}
\providecommand{\BIBentryALTinterwordstretchfactor}{4}
\providecommand{\BIBentryALTinterwordspacing}{\spaceskip=\fontdimen2\font plus
\BIBentryALTinterwordstretchfactor\fontdimen3\font minus
  \fontdimen4\font\relax}
\providecommand{\BIBforeignlanguage}[2]{{%
\expandafter\ifx\csname l@#1\endcsname\relax
\typeout{** WARNING: IEEEtran.bst: No hyphenation pattern has been}%
\typeout{** loaded for the language `#1'. Using the pattern for}%
\typeout{** the default language instead.}%
\else
\language=\csname l@#1\endcsname
\fi
#2}}
\providecommand{\BIBdecl}{\relax}
\BIBdecl

\bibitem{VanEyk2018}
E.~van Eyk, L.~Toader, S.~Talluri, L.~Versluis, A.~Uta, and A.~Iosup,
  ``{Serverless is More: From PaaS to Present Cloud Computing},'' \emph{IEEE
  Internet Computing}, vol.~22, no.~5, pp. 8--17, sep 2018.

\bibitem{Castro2019}
P.~Castro, V.~Ishakian, V.~Muthusamy, and A.~Slominski, ``{The rise of
  serverless computing},'' \emph{Communications of the ACM}, vol.~62, no.~12,
  pp. 44--54, 2019.

\bibitem{Morabito2017}
R.~Morabito, ``{Virtualization on internet of things edge devices with
  container technologies: A performance evaluation},'' \emph{IEEE Access},
  vol.~5, pp. 8835--8850, 2017.

\bibitem{Santos2019}
J.~Santos, T.~Wauters, B.~Volckaert, and F.~{De Turck}, ``{Towards
  network-Aware resource provisioning in kubernetes for fog computing
  applications},'' \emph{Proceedings of the 2019 IEEE Conference on Network
  Softwarization: Unleashing the Power of Network Softwarization, NetSoft
  2019}, pp. 351--359, 2019.

\bibitem{Adzic2017}
G.~Adzic and R.~Chatley, ``{Serverless computing: economic and architectural
  impact},'' in \emph{Proceedings of the 2017 11th Joint Meeting on Foundations
  of Software Engineering}.\hskip 1em plus 0.5em minus 0.4em\relax {ACM}, aug
  2017, pp. 884--889.

\bibitem{Carpio2020}
F.~Carpio, M.~Delgado, and A.~Jukan, ``{Engineering and Experimentally
  Benchmarking a Container-based Edge Computing System},'' in \emph{{ICC} 2020
  {IEEE} International Conference on Communications ({ICC})}.\hskip 1em plus
  0.5em minus 0.4em\relax {IEEE}, 2020.

\bibitem{Li2017a}
C.~Li, Y.~Xue, J.~Wang, W.~Zhang, and T.~Li, ``{Edge-Oriented Computing
  Paradigms : A Survey on Architecture},'' \emph{{ACM} Computing Surveys},
  vol.~xx, no.~x, pp. 1--34, jun 2017.

\bibitem{Wang2020}
X.~Wang, Y.~Han, V.~C. Leung, D.~Niyato, X.~Yan, and X.~Chen, ``{Convergence of
  Edge Computing and Deep Learning: A Comprehensive Survey},'' \emph{IEEE
  Communications Surveys and Tutorials}, vol.~22, no.~2, pp. 869--904, 2020.

\bibitem{Das2019a}
A.~Das, S.~Patterson, and M.~Wittie, ``{EdgeBench: Benchmarking Edge Computing
  Platforms},'' in \emph{2018 IEEE/ACM International Conference on Utility and
  Cloud Computing Companion (UCC Companion)}.\hskip 1em plus 0.5em minus
  0.4em\relax IEEE, dec 2018, pp. 175--180.

\bibitem{Caprolu2019}
M.~Caprolu, R.~{Di Pietro}, F.~Lombardi, and S.~Raponi, ``{Edge Computing
  Perspectives: Architectures, Technologies, and Open Security Issues},''
  \emph{2019 IEEE International Conference on Edge Computing (EDGE)}, pp.
  116--123, 2019.

\bibitem{Morabito2018}
R.~Morabito, V.~Cozzolino, A.~Y. Ding, N.~Beijar, and J.~Ott, ``{Consolidate
  IoT Edge Computing with Lightweight Virtualization},'' \emph{IEEE Network},
  vol.~32, no.~1, pp. 102--111, 2018.

\bibitem{Loghin2019}
D.~Loghin, L.~Ramapantulu, and Y.~M. Teo, ``{Towards Analyzing the Performance
  of Hybrid Edge-Cloud Processing},'' \emph{2019 IEEE International Conference
  on Edge Computing (EDGE)}, no.~i, pp. 87--94, 2019.

\bibitem{Palade2019}
A.~Palade, A.~Kazmi, and S.~Clarke, ``{An Evaluation of Open Source Serverless
  Computing Frameworks Support at the Edge},'' \emph{2019 IEEE World Congress
  on Services (SERVICES)}, vol. 2642-939X, pp. 206--211, 2019.

\bibitem{loria_2020_sloriatextblob}
\BIBentryALTinterwordspacing
S.~Loria, ``sloria/textblob,'' GitHub, 04 2020. [Online]. Available:
  \url{https://github.com/sloria/textblob}
\BIBentrySTDinterwordspacing

\bibitem{Szegedy_2016}
C.~Szegedy, V.~Vanhoucke, S.~Ioffe, J.~Shlens, and Z.~Wojna, ``Rethinking the
  inception architecture for computer vision,'' jun 2016.

\end{thebibliography}

\end{document}